\documentclass[aip,graphicx]{revtex4-1}
\usepackage[version=3]{mhchem}
\usepackage{graphicx}

\draft 

\begin{document}


\title{Natural band alignment of $\rm MgO_{1-x}S_{x}$ alloys} 



\author{Yuichi Ota}
\affiliation{Tokyo Metropolitan Industrial Technology Research Institute,2-4-10 Aomi, Koto-ku, Tokyo 135-0064, Japan}
\email[]{ota.yuichi@iri-tokyo.jp}
\author{Kentaro Kaneko}
\affiliation{Research Organization of Science and Technology, Ritsumeikan University, Kusatsu 525-8577, Japan}

\author{Takeyoshi Onuma}%
\affiliation{Department of Applied Physics, School of Advanced Engineering and Department of Electrical Engineering and Electronics, Graduate School of Engineering, Kogakuin University, 2665-1 Nakano, Hachioji, Tokyo 192-0015, Japan}%

\author{Shizuo Fujita}
\affiliation{Office of Society-Academia Collaboration for Innovation, Kyoto University, Kyoto 606-8501, Japan}%


\date{\today}

\begin{abstract}
We have calculated formation enthalpies, band gaps, and natural band alignment for $\rm MgO_{1-x}S_{x}$ alloys by first principles calculation based on density functional theory.
The calculated formation enthalpies show that the $\rm MgO_{1-x}S_{x}$ alloys exhibit a large miscibilitygap, and a metastable region was found to occur when the S content was below 18\% or over 87\%.
Effect of S incorporation for band gaps of  $\rm MgO_{1-x}S_{x}$ alloys shows large bowing parameter (b $ \simeq $ 13 eV) induced.
The dependence of the band lineup of $\rm MgO_{1-x}S_{x}$ alloys on the S content by using two different methods, and the change in the energy position of valence band maximum (VBM) was larger than that of conduction band minimum. 
Based on the calculated VBM positions, we predicted that $\rm MgO_{1-x}S_{x}$ with S content 10 to 18\% can be surface charge transfer doping by high electron affinity materials.
The present work provides an example to design for p-type oxysulfide materials. 
\end{abstract}


\maketitle 

\section{INTRODUCTION}
Magnesium oxide (MgO) is a non-toxic oxide material commonly used in refractory and medical applications.
It has a wide bandgap ($\sim$7.8 eV) and direct transition type electronic band structure, which is a promising material for optical device application\cite{onuma2021identification}.
Despite these attractive physical properties, MgO has only been utilized as a barrier material in electronics applications such as spin-tunneling devices\cite{parkin2004giant}.
This is probably because MgO is considered to be carrier uncontrollable and cannot be used as a semiconductor.
However, MgO exhibits a standard semiconducting mechanism where Li doping increases electrical conductivity\cite{tardio2002p}.
Furthermore,  in recent years, bandgap engineering has been demonstrated by epitaxial growth of rock-salt (RS) structured MgZnO alloys\cite{kaneko2016growth,kaneko2018deep,onuma2018impact,ishii2019pure,ono2019excitation,gorczyca2020rocksalt}.
Therefore, MgO and MgZnO alloys are attracting attention as new ultra-wide bandgap (UWBG) semiconductors\cite{tsao2018ultrawide,higashiwaki2021ultrawide}. 

 Among the known UWBG semiconductors, oxide materials such as \ce{Ga2O3} and \ce{Al2O3} in particular are known to have difficulty with p-type (or ambipolar) doping\cite{yan2008doping}.
According to the modern theory of doping, some UWBG materials (also MgO) are unable to control their Fermi energy to near the valence band maximum (VBM) or conduction band minimum (CBM)\cite{cao2019design,goyal2020dopability,zunger2021understanding}.
This dopability trend has been empirically explained by the location of the band edges (i.e., VBM and CBM) relative to the vacuum level ($E_{vac}$).
The empirical criterion indicates that holes can be doped when the VBM is greater than approximately $-6$ eV, meaning that a shallower VBM is advantageous for p-type doping in oxide materials\cite{robertson2011limits,hosono2013exploring,robertson2021doping}.
P-type wide bandgap oxides such as \ce{NiO}\cite{sawatzky1984magnitude}, \ce{CuAlO2}\cite{yanagi2000electronic}, \ce{SrCu2O2}\cite{ohta2002electronic}, and $\alpha$-\ce{Ir2O3}\cite{kan2018electrical} have been reported hybridized orbitals between O 2p and transition metal d produce shallower VBM than other oxides.
Transition metal oxides were found to be suitable for p-type doping, but it is difficult to obtain a UWBG (\textgreater \: 4 eV) using these compounds due to the influence of d electrons with low binding energy.
Despite this trade-off, the alloying of $\alpha$-\ce{Ga2O3} and $\alpha$-\ce{Ir2O3} has successfully realized p-type UWBG oxides\cite{kaneko2021ultra,kaneko2022novel}.
In the case of VBM for $\alpha$-(Ir,Ga)$_2$\ce{O3}, the energy positions depend on the Ga (Ir) atom compositions\cite{kaneko2021ultra}.
Thus, attempts to search for p-type UWBG oxides has focused on those containing cation atoms with d-bands.

 Another approach to p-type doping involves alloying the semiconductor to raise the host VBM toward $E_{vac}$ and decrease the acceptor level relative to it\cite{cao2019design}.
For example, it was found that when zinc oxide (ZnO) is alloyed with the effect of sulfur (S), the nitrogen ($\rm N_{O}$) acceptor ionization energy is reduced for pure ZnO\cite{persson2006strong}.
These results indicated that a large VBM shift occurs when the oxygen that predominantly forms the VBM of ZnO is replaced with another anion atom (S). 
Therefore, the VBM of oxide semiconductors can be better modulated by substituting anion atoms rather than cation atoms. 
Unfortunately, only a small number of experimental investigations have been conducted on doped p-type $\rm ZnO_{1-x}S_{x}$ because of the strong unintentional n-type doping of ZnO\cite{kang2014structural,kobayashi2017growth}. 
Thus the well-established studies of cation-substituted II-VI semiconductor alloys, there are few reports on anion-substituted ones\cite{adachi2009properties}.
Therefore, the common-cation systems such as $\rm MgO_{1-x}S_{x}$ alloys should be investigated not only for p-type doping of UWBG materials, but also for future oxysulfide (or oxychalcogenides) applications\cite{woods2020wide}.

In this paper, we address to analyze the effect of sulfur incorporation on MgO by using first principles calculation based on density functional theory (DFT).
The formation enthalpies and band gap of $\rm MgO_{1-x}S_{x}$ alloys were investigated, and the natural band alignment of the alloys was also derived from the calculations to predict doping tendency.
Moreover, we proposed to combine $\rm MgO_{1-x}S_{x}$ alloys with high electron affinity materials that could induce hole carriers by surface charge transfer doping (SCTD).

\section{CALCULATION DETAILS}
We modeled the supercell structure of the $\rm MgO_{1-x}S_{x}$ alloy; the alloy cells contained 16 atoms, namely 8 Mg and 8 O (S) atoms. 
To consider only inequivalent cells due to the atomic substitution, we employed a \textit{supercell program}\cite{okhotnikov2016supercell}. 
The geometry optimization was performed using the Quantum ESPRESSO package\cite{giannozzi2009quantum,giannozzi2017advanced}. 
We used the Perdew-Burke-Ernzerhof form within the generalized gradient approximation with projector-augmented wave potentials\cite{perdew1996generalized,blochl1994projector}.
The electronic wavefunctions and charge densities were expanded in plane waves with cut-off energies of 65 and 450 Ry, respectively. 
The \textit{k}-points were generated using the Monkhorst-Pack scheme with a mesh size of 8 $\times$ 8 $\times$ 8 for optimizing the structure calculations\cite{monkhorst1976special}.
To optimize the model geometry, the stress and force on the atoms were set to 0.05 GPa and 0.1 mRy/a.u., respectively. 
After optimizing the model geometry, the most favorable alloy structures were determined using the formation enthalpy. 
The formation enthalpy, $\Delta H$, was calculated according to

\begin{equation}
	\begin{split}
		\label{eq0}
		\Delta H\left[ \rm MgO_{1-x}S_{x} \right]  = E\left[\rm MgO_{1-x}S_{x} \right] 
		- \left( 1-x\right) E\left[\ce{MgO}\right] - xE\left[\ce{MgS}\right],
	\end{split}
\end{equation}

where E$\left[\ce{MgO}\right]$ and E$\left[\ce{MgS}\right]$ are the energies of the lowest RS structures. 
A regular solution model for alloy energies can be described as

\begin{equation}
	\label{eq01}
	\Delta H\left[ \rm MgO_{1-x}S_{x} \right]  = 4x(1-x)\Delta H_{0},
\end{equation}

and $\Delta H_{0}$ represented critical formation enthalpy\cite{neugebauer1995electronic}.
The growth temperature prediction using following equation
\begin{equation}
	\label{eq02}
	T  = \Delta H_{0}/k_{b}*(8x-4)/[\ln x - \ln(1-x)],
\end{equation}

where $k_{b}$ is the Boltzmann constant\cite{neugebauer1995electronic}.
All favorable alloy structures were considered to calculate the electronic band structures using the Wien2k code\cite{blaha2020wien2k}.
The calculation details are provided as supplementary information. 
To determine the natural band edge positions of the alloys using the two-alignment method, we employed a modified Tersoff method\cite{schleife2009branch} for the branch point energy ($E_{BP}$) and the atomic solid-state energy (SSE) scale approach\cite{ota2020band}.
Originally, $E_{BP}$ means mid gap energy of semiconductor, which is given by averaged valence and conduction band energies.
The $E_{BP}$ is here approximated as

\begin{equation}
	\label{eq03}
	E_{BP} \approx \frac{1}{2N_\mathbf{k}} \sum_{\mathbf{k}} \left[  \frac{1}{N_{CB}}\sum_{i}^{N_{CB}} \epsilon_{c_i}(\mathbf{k}) +  \frac{1}{N_{VB}}\sum_{j}^{N_{VB}} \epsilon_{v_j}(\mathbf{k}) \right], 
\end{equation}

where $N_{\mathbf{k}}$ is the number of $\mathbf{k}$ points in the $\mathbf{k}$ meshes in the Brillouin zone, and $\epsilon_{c_i}$ and $\epsilon_{v_j}$ are the \textit{i}th lowest conduction band and \textit{j}th highest valence band states at the wave vector \textit{k}, respectively.
We calculated $E_{BP}$ using 8 $N_{CB}$ and 16 $N_{VB}$ for the alloy structures, and 2 $N_{CB}$ and 1 $N_{VB}$ for the primitive structures, such as MgO and MgS. 
An atomic SSE was determined by an averaged electron affinity (for a cation atom) or an ionization potential (for an anion atom) for several inorganic compounds\cite{pelatt2011atomic}.
The SSE scale approach can be determined natural band alignment using the SSE and band gap of material\cite{ota2020band}. 
1.72, 7.98, and 6.31 eV were used as the absolute values of the atomic SSEs of Mg, O, and S, respectively\cite{pelatt2015atomic}.

\section{Results and Discussions}
\begin{figure}[p!]
	\includegraphics[scale=0.55]{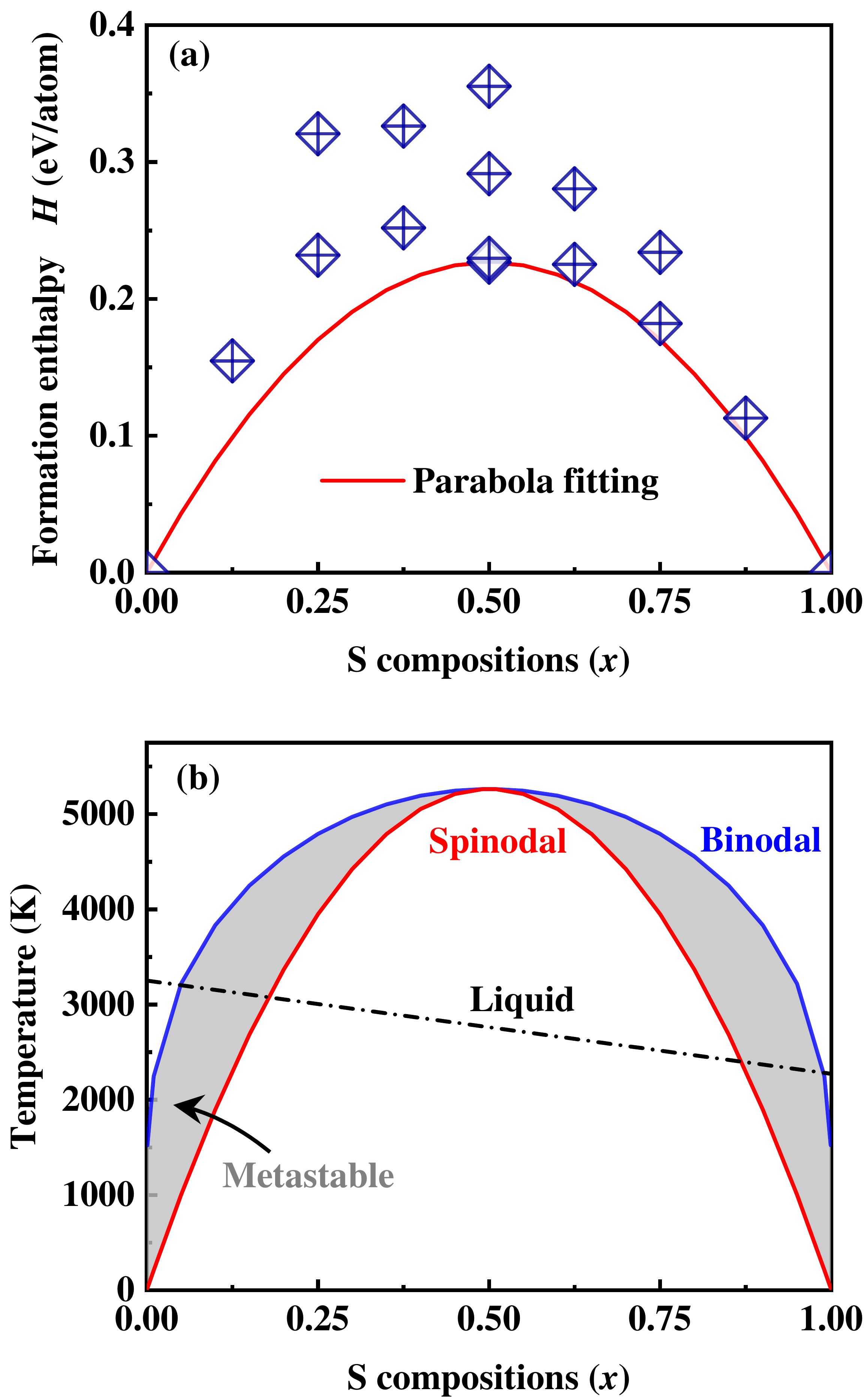}
	\caption{ (a) Enthalpy of formation per atom as a function of S content for rock-salt (RS) structured $\rm MgO_{1-x}S_{x}$ alloys with parabola fitting by Eq \ref{eq01}.
		(b) Lower limit of the miscibility gap for $\rm MgO_{1-x}S_{x}$ alloys. The liquidus line shows the separation of the liquid from the solid phase.}
	\label{fgr:enthalpy}
\end{figure}

Figure \ref{fgr:enthalpy} (a) shows the formation enthalpy per atom of $\rm MgO_{1-x}S_{x}$ alloys based on the use of all nonequivalent atomic substitution models. In this calculation, we did not consider the wurtzite structure or the alloy mixing enthalpy.
We have estimated critical formation enthalpy ($\Delta H_{0}$ = 227 meV) by equation \ref{eq01} parabola fitting.
The growth temperatures of $\rm MgO_{1-x}S_{x}$ alloys using $\Delta H_{0}$ and equation \ref{eq02} are depicted in Fig.\ref{fgr:enthalpy} (b).
These results show  $\rm MgO_{1-x}S_{x}$ alloys single crystal growth is difficult in thermal equilibrium conditions\cite{millican2020alloying}.
However, metastable regions exist in S concentrations less than 18\% and S more than 87\% below the liquidus line.
Therefore, we believe that the crystal growth of $\rm MgO_{1-x}S_{x}$ alloys can be achieved by the mist chemical vapor deposition (CVD) method, which is suitable for the growth of oxide metastable structures and their alloys\cite{shinohara2008heteroepitaxy,ito2012growth,fujita2014epitaxial, suzuki2014growth,kaneko2016growth2,fujita2016evolution,oshima2017epitaxial,dang2018bandgap,jinno2018control,tahara2018heteroepitaxial,nishinaka2018incorporation,horie2021epitaxial,takane2021initial,ogura2022alloying,biswas2022thermodynamically,nishinaka2022growth,kaneko2022prospects,takane2022band}. 
Indeed, mist CVD has succeeded in synthesizing sulfides such as \ce{ZnS}\cite{uno2016growth,uno2016growth2,uno2017photoluminescence,okita2018structural} and \ce{Cu2SnS3}\cite{okamura2022fabrication}, and the growth of metastable $\rm MgO_{1-x}S_{x}$ alloys is a future challenge. 
Note that the S compositions outside of metastable regions in Fig.\ref{fgr:enthalpy} (b), $\rm MgO_{1-x}S_{x}$ alloys may grow in an amorphous phase\cite{jaquez2019amorphous}.

\begin{figure}[p!]
	\includegraphics[scale=0.6]{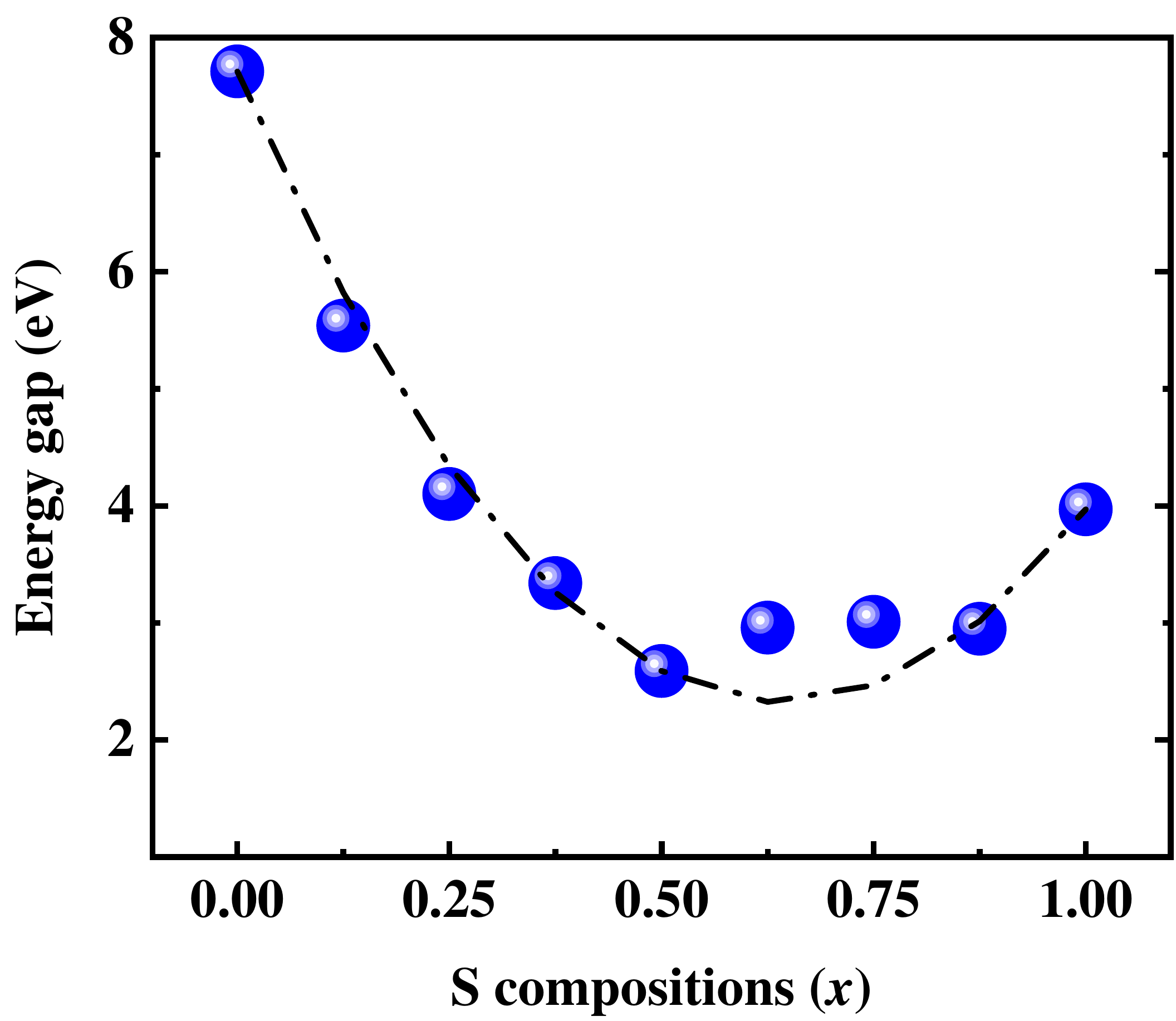}
	\caption{ Energy gaps of $\rm MgO_{1-x}S_{x}$ alloys as a function of S composition.}
	\label{fgr:bandgap}
\end{figure}

To estimate the bandgap ($E_{g}$) of $\rm MgO_{1-x}S_{x}$ alloys, we calculated the lowest $\Delta H$ models using the Wien2k code.
The result is shown in Figure \ref{fgr:bandgap}. The $E_{g}$ dependence on the S content of the $\rm MgO_{1-x}S_{x}$ alloy was determined using the Tran–Blaha modified Becke Johnson potential\cite{tran2009accurate,koller2012improving}.
The calculated trend of $E_{g}$ as a function of the S content exhibits the same nonlinear relationship behavior as that experimentally observed for wurtzite structured $\rm ZnO_{1-x}S_{x}$ alloys\cite{meyer2004structural}.
The $E_{g}$ of  $\rm MgO_{1-x}S_{x}$ alloys is described as

\begin{equation}
	\label{eq04}
	E_{g} (x) = (1-x)E_{g}[\ce{MgO}]  + xE_{g}[\ce{MgS}]-bx(1-x),
\end{equation}

where $b$ is the bowing parameter, and the bandgap energies $E_{g}$[MgO] and $E_{g}$[MgS] are 7.71 and 3.91 eV, respectively.
We estimated that the bowing parameter is approximately 13 eV due to the large difference in the $E_{g}$ values between MgO and MgS.
This large bowing parameter is similar to that of corundum structured $\rm Al_{2}(O_{1-x}Se_{x})_{3}$ (b $ \simeq $ 19 eV)\cite{liu2021large}, wurtzite structured $\rm GaN_{1-x}As_{x}$ (b $ \simeq $ 16.9 eV)\cite{kimura2004epitaxial}, $\rm AlN_{1-x}P_{x}$ (b $ \simeq $ 28.3 eV)\cite{borovac2018first}, and $\rm AlN_{1-x}As_{x}$ (b $ \simeq $ 30.5 eV)\cite{tan2016dilute} structures, and it is thought that large bowing parameters are a unique feature of highly mismatched alloys with a large difference in $E_{g}$.
Note that the $E_{g}$ of MgO agrees well with the experimental value.
By contrast, the $E_{g}$ of MgS is difficult to compare with the calculation results because there are few experimental reports for this material.
However, our calculated $E_{g}$ value of 3.91 eV seems reasonable for the RS-structured MgS since previous theoretical calculations have reported values ranging between 3.2 and 4.8 eV\cite{bhandari2018predictions}, and experimental studies have measured values ranging between 3.5 and 4.1 eV\cite{taleatu2015microstructural}. 
Therefore, our calculated $E_{g}$ values of the $\rm MgO_{1-x}S_{x}$ alloys are suitable for band alignment estimation.

\begin{figure}[p!]
	\includegraphics[scale=0.6]{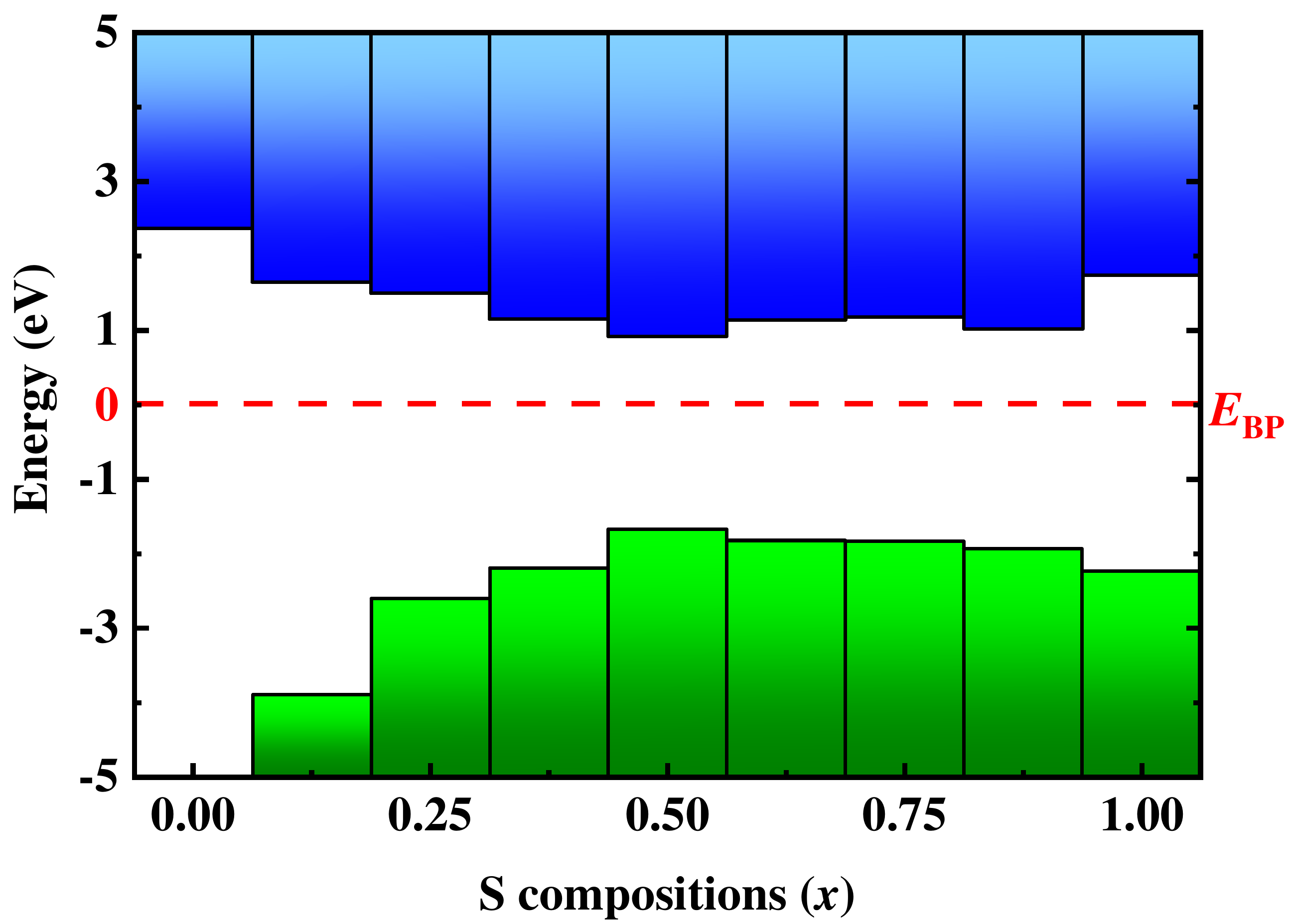}
	\caption{Band edges for $\rm MgO_{1-x}S_{x}$ alloys aligned to branch point energy ($E_{BP}$). The lower line indicated the valence band maximum (VBM) and upper line the positions of the conduction band minimum (CBM).}
	\label{fgr:bpe}
\end{figure}

\begin{figure}[p!]
	\centering
	\includegraphics[scale=0.6]{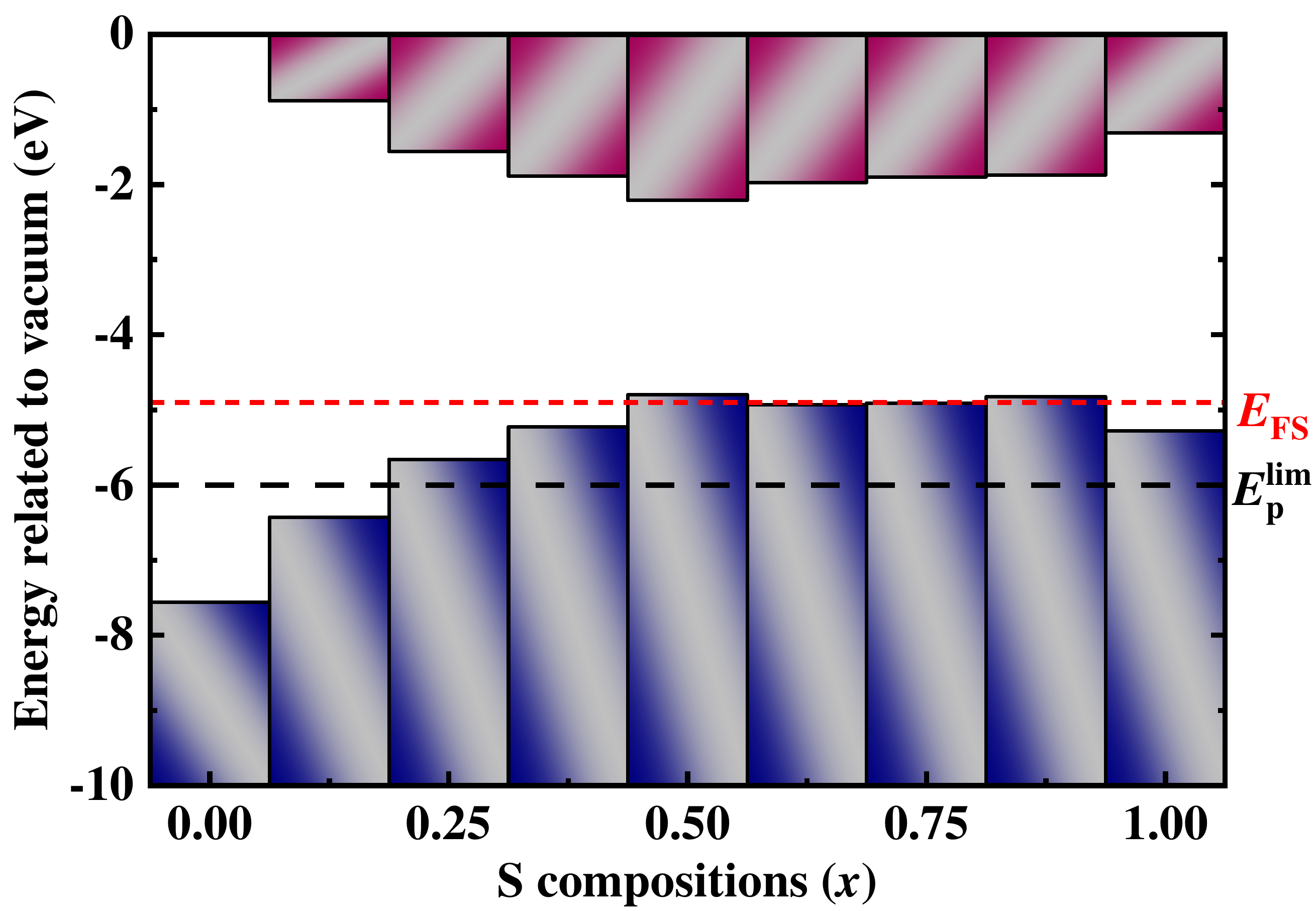}
	\caption{Band alignment relative to the vacuum level for $\rm MgO_{1-x}S_{x}$ alloys via atomic solid-state energy (SSE) scale approach. }
	\label{fgr:sse}
\end{figure}

The resulting natural band lineups to $E_{BP}$ are plotted in Figure \ref{fgr:bpe}; they are aligned with respect to the red dotted line. 
The $E_{BP}$ represents a charge neutrality level (CNL), which marks the energy where defect states change their character from predominantly donor-like (acceptor-like).
Unintentional heavy p-type doping is expected to occur when the $E_{BP}$ line overlaps or below the VBM\cite{sarmadian2016easily}. 
This shows that the energy formation of acceptor-type defects becomes smaller than that of donor-type defects\cite{king2009unification}.
According to this simple classification concept, we can infer that the closer the $E_{BP}$ is to the VBM of the alloys, the higher the potential will be for p-type doping.
The lowest $E_{BP}$ (1.67 eV) occurred for the $\rm MgO_{0.5}S_{0.5}$ case; this value is significantly lower than that of MgO ($E_{BP}$ = 5.34 eV). 
However, the tunability of the VBM position with the S content is lost at approximately 2 eV below the $E_{BP}$. 
The natural band lineup result indicates that S incorporation enhances the possibility of p-type doping, although the modified Tersoff method does not provide band edge positions on an absolute energy scale. 
Therefore, we employed the atomic SSE approach to evaluate the band discontinuities of $\rm MgO_{1-x}S_{x}$ alloys. 
The natural band alignment relative to the $E_{vac}$ of the $\rm MgO_{1-x}S_{x}$ alloys is illustrated in Figure \ref{fgr:sse}. 
The different trends of each band edge position as a function of the S content were found to be consistent with Figure \ref{fgr:bpe}. 
Here we estimated the possibility of achieving p-type doping of $\rm MgO_{1-x}S_{x}$ alloys using several doping limitation indicators. 
Some works have presented universal CNLs\cite{ota2022natural}, such as the hydrogen $\epsilon (+/-)$  level\cite{van2003universal}, limiting Fermi level ($F_{lim}$)\cite{brudnyui2007model}, and Fermi level stabilization energy ($E_{FS}$)\cite{walukiewicz2001intrinsic}. 
The $\epsilon (+/-)$ level, $F_{lim}$, and $E_{FS}$ are located at about $-4.5$, $-4.7 \pm 0.2$\cite{brudnyi2004model}, and $-4.9$ eV below $E_{vac}$, respectively.
These CNLs are indicators used to discriminate whether extrinsic defects or hydrogen are donors or acceptors depending on the Fermi level in a material. 
However, in this work, we focus only on the deepest $E_{FS}$ level from $E_{vac}$.
Moreover, following the empirical rule, hole doping of oxide is possible when the VBM is located above approximately $-6$ eV\cite{robertson2011limits,hosono2013exploring,robertson2021doping}.
These different indicators originate from different approaches; they all seem to indicate the intrinsic doping criteria for semiconductors.
According to the empirical rule ($E_{lim}^p$), our calculated band lineup of the $\rm MgO_{1-x}S_{x}$ alloys shows that  hole doping becomes possible when $x$ exceeds 0.25.
Such a trend is understandable given the $E_{FS}$ using the amphoteric native defect model\cite{walukiewicz2001intrinsic}.
As shown in Figure \ref{fgr:sse}, the VBM of $\rm MgO_{1-x}S_{x}$ (0.5 $\leq$ x $\leq$ 0.875) is located almost at the $E_{FS}$, which explains the strong possibility for p-type doping at these S compositions. 
Since the smaller the energy difference $|E_{FS}-VBM|$, the larger the maximum hole carrier concentration, $\rm MgO_{1-x}S_{x}$ alloys are expected to exhibit enhanced p-type conduction with respect to MgO\cite{tokumitsu1990correlation}.
Therefore, the empirical rule and the $E_{FS}$ trends are consistent as p-type dopability indicators. 
We note that the finding suitable dopants for an alloy is difficult with the current DFT calculation framework.

Although Figure \ref{fgr:sse} shows a trend similar to the VBM change obtained via the $E_{BP}$ result, it reveals a larger hole-doping possibility than Figure \ref{fgr:bpe}. 
Note that in several recent studies, the modified Tersoff method ($E_{BP}$) has not correctly predicted the p-type dopability trend\cite{yim2018computational,woods2018assessing,brunin2019transparent}. 
Consequently, we expect the $E_{FS}$ value and the empirical rule ($E_{lim}^p$) to be better predictors of p-type doping criteria. 
These band alignment results indicate that doping MgO with both electrons and holes is extremely difficult.
For example, lithium (Li) is a deep acceptor level in MgO, and its activation energy has been reported to be about 0.7 eV\cite{tardio2002p}.
However, the Li acceptor activation energy is expected to decrease as the VBM moves upward upon S alloying\cite{goodrich2021prospects}. 
Furthermore, since the defect and impurity levels are determined by the host material\cite{huang2015alloy}, S concentrations of about 25\% or less for $\rm MgO_{1-x}S_{x}$ alloys can be expected to lower the Li acceptor levels by increasing the energy position of the VBM, as in the case of $\rm ZnO_{1-x}S_{x}$ alloys\cite{persson2006strong}. 
Additionally, universal levels of oxygen\cite{chakrapani2021universal} and transition metal impurities\cite{caldas1984universal,zunger1985theory} have been reported, and these could become shallow acceptor levels through VBM modulation. 
This experimental evidence is not yet available; however, a theoretical result based on defect formation energy calculations suggested that MgS is a p-type transparent conducting material candidate\cite{raghupathy2018rational,li2019first}.
This result is consistent with those based on the empirical rule and the $E_{FS}$ predictions, and we thus believe that our calculations can correctly predict the p-type dopability criteria.

\begin{figure}[p!]
	\centering
	\includegraphics[scale=0.7]{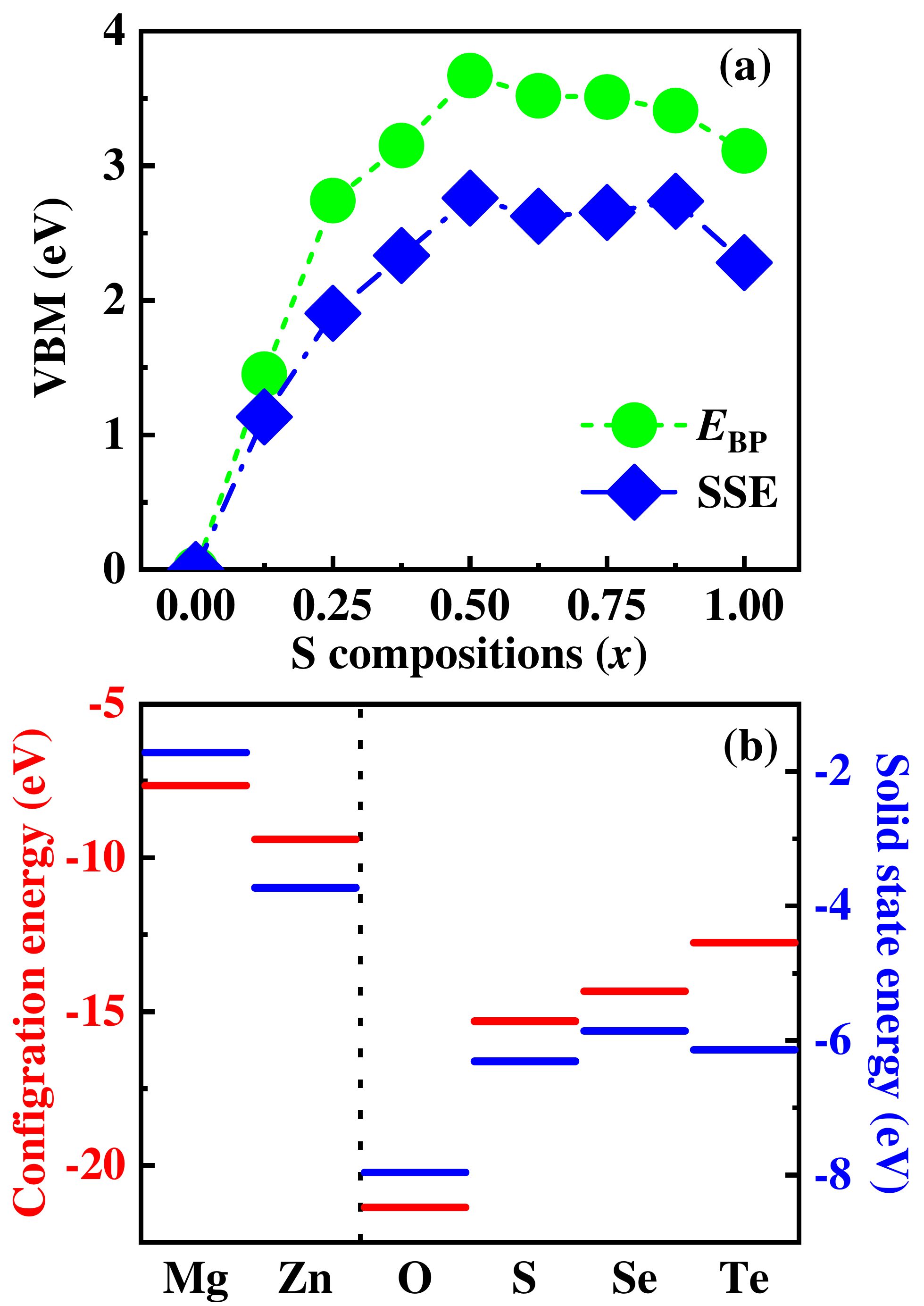}
	\caption{(a) Relative positions of valence band maximum (VBM) with respect to magnesium oxide (MgO) by branch point energy ($E_{BP}$) and solid-state energy (SSE) method. (b) The locations of the II-VI atomic energy level determined by configuration energies (CEs) and SSE. }
	\label{fgr:CE&SSE}
\end{figure}

The calculated relative VBM positions ($\Delta$VBM) concerning MgO are shown in Figure \ref{fgr:CE&SSE} (a). 
These results indicate that the increasing trends of the natural valence band offset of $\rm MgO_{1-x}S_{x}$ with increasing S content obtained using different approaches are consistent. 
The offset value trend can be explained as a chemical trend by considering the atomic energy of common anions, such as O, S, Se, and Te atoms. 
Figure \ref{fgr:CE&SSE} (b) shows the atomic energy levels, including the configuration energy (CE)\cite{mann2000configuration} and SSE\cite{pelatt2015atomic} for each atom. 
The energies of these VI-group atoms become closer to $E_{vac}$ as the atomic number increases, except for the SSE of Te atoms. 
CE refers to the energy of isolated atoms, whereas SSE refers to the energy of \textit{“atoms in a crystal”}, such as the Bader charge concept\cite{bader1985atoms}. 
Our calculation results can be explained by the SSE more simply than those derived from the CE.
For example, a rough estimate of $E_{g}$ can be obtained by calculating the difference in SSE between cation atoms, such as Mg and Zn, and anion atoms (cf. Figure \ref{fgr:CE&SSE} (b))\cite{pelatt2011atomic}.
The VBM of $\rm MgO_{1-x}S_{x}$ can be determined by calculating the difference in the SSE scale between S and O atoms. 
These findings indicate that the energy levels of atoms in a crystal are determined by the oxidation (or reduction) state\cite{pelatt2019elucidation}. 
Therefore, the energy position of the VBM of ionic bonded materials can be modulated via anion atom substitutions. 
We note that the VBM modulation is only possible when anion atoms dominate the top of the valence band of the material (see Supporting information Figures S1-S9). 
These results clearly show that Mg atoms contribute little to the valence band in the energy range of $-5$ $\sim$ 0 eV.

\begin{figure}[p!]
	\centering
	\includegraphics[scale=0.7]{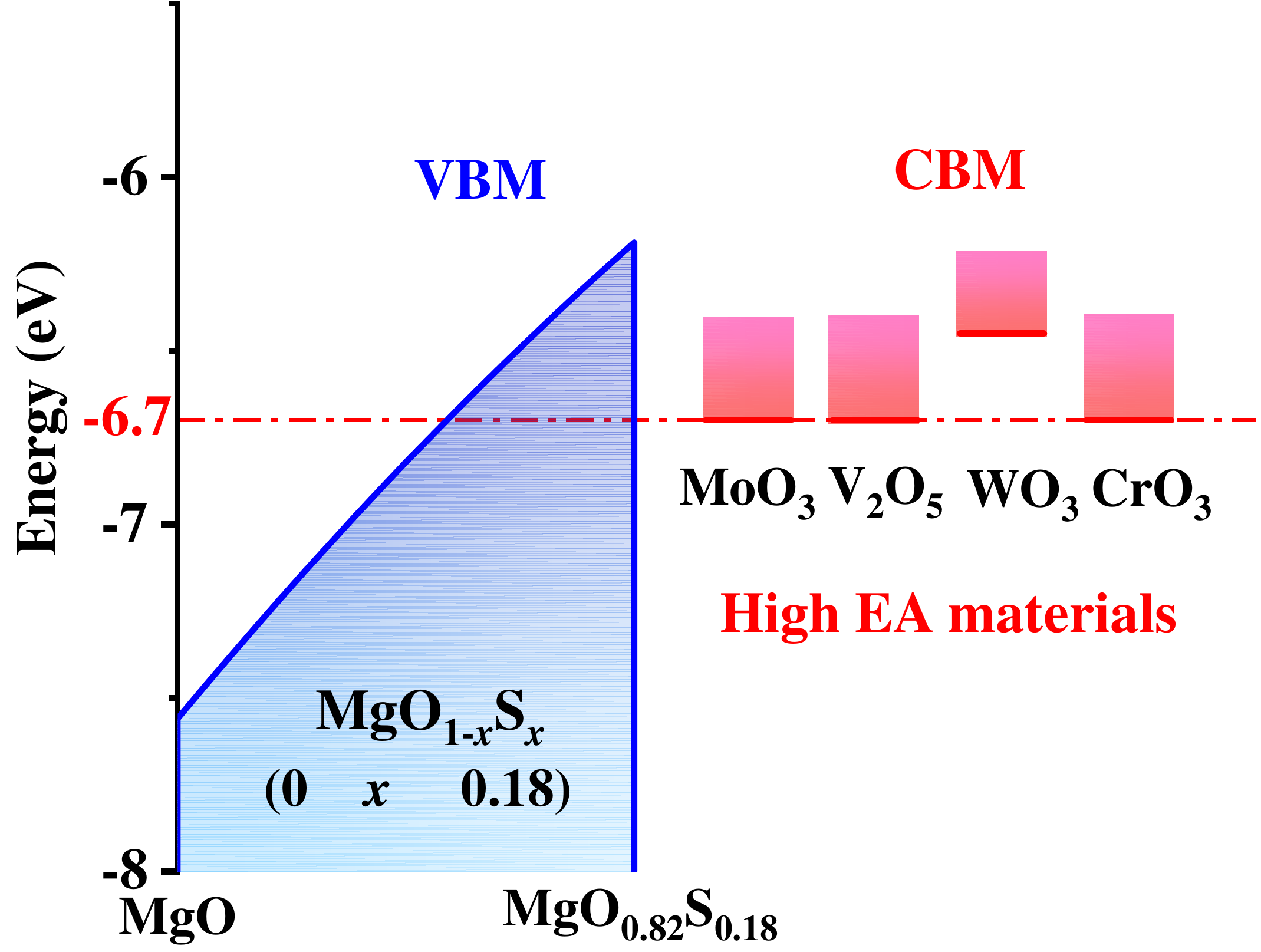}
	\caption{Schematic drawing the VBM positions for $\rm MgO_{1-x}S_{x}$ alloys with CBM positions for high electron affinity materials.  }
	\label{fgr:SCTD}
\end{figure}

 From our band alignment results, we have considered hole doping by SCTD\cite{chen2009surface}.
The SCTD technique has been demonstrated for hydrogenated diamonds\cite{crawford2021surface} and several inorganic semiconductors : graphene\cite{chen2007surface}, silicon\cite{rietwyk2014charge}, cubic boron nitride\cite{he2015surface}, and perovskite materials\cite{euvrard2021electrical}.
In order to holes doping spontaneously, we have to contact the high electron affinity (EA) or high work function materials with $\rm MgO_{1-x}S_{x}$ alloys.
Here we have investigated whether SCTD occurs for metastable $\rm MgO_{1-x}S_{x}$ condition (x $\leq$ 0.18) with $E_{g}$ over 4 eV shown in Fig \ref{fgr:SCTD}.
The high EA materials are approximately 6.7 eV for \ce{MoO3}\cite{kroger2009role}, \ce{V2O5}\cite{meyer2011electronic}, \ce{CrO3}\cite{greiner2012universal}, and approximately 6.45 eV for \ce{WO3}\cite{meyer2010charge}, respectively.
The results indicated that the surface charge transfer appeared spontaneously at S content above 10\% for $\rm MgO_{1-x}S_{x}$ alloys in contact with high EA materials except for \ce{WO3}.
Note that the energetic positions of the VBM for $\rm MgO_{1-x}S_{x}$ alloys are calculated by the SSE scale approach, while the CBMs for high EA materials are experimental values.
These high EA materials induced sheet hole carrier concentrations greater than $10^{13}$ cm$^{-2}$ with respect to hydrogenated diamond\cite{verona2016comparative,crawford2021surface}.
Thus, the sheet hole carrier concentration of $\rm MgO_{1-x}S_{x}$ alloys is expected to be the same order of magnitude hydrogenated diamond by SCTD.
This is consistent with the number of adsorbed molecules on the MgO (100) or (111) surface\cite{onishi1987adsorption}, and the sheet hole carriers are expected to be tunable around $10^{14}$ cm$^{-2}$.
Other potential surface charge-induced phenomena include defect modulation doping\cite{weidner2019defect}, formal (or surface) polarization effects\cite{zhou2015surface,adamski2020polarization}, and electrochemical doping \cite{ishii1999energy,wang2017co,rietwyk2019universal,wang2019role}.
These phenomena are different from \textit{bulk doping} but show that holes can be generated by using appropriate surface or interface effects.
Therefore, these SCTD schemes must be adopted instead of conventional impurity doping to realize p-type UWBG oxide semiconductors.

\section{Conclusion}
We have investigated the S incorporation effect of MgO using a first principles calculation based on DFT.
Our results indicated that the solubility of S content required for the metastable solid phase of the $\rm MgO_{1-x}S_{x}$ alloys was estimated to be less than 18\% or over 87\%, respectively.
The band gap of $\rm MgO_{1-x}S_{x}$ alloys became the smallest at 50\% S content, and the bowing parameter was estimated b $ \simeq $ 13 eV.
Our calculated natural band alignment indicated that the $\rm MgO_{1-x}S_{x}$ alloys tend to be more easily doped to p-type than n-type.
Based on the calculated VBM positions, we predicted that metastable $\rm MgO_{1-x}S_{x}$ with S content of 10 to 18\% can be SCTD by high EA materials.
We hope that the present work may provide design guidelines for p-type oxysulfide materials.

\begin{acknowledgments}
This work was supported in part by Grants-in-Aid for Scientific Research No. 20H00246 from MEXT, Japan.
\end{acknowledgments}

\section*{AUTHOR DECLARATIONS}
\section*{Conflict of Interest}
The authors have no conflicts to disclose.

\section*{Supplementary Material}
Details of the DFT calculations and their results are provided in the supplementary materials.

\section*{Author Contributions}
\textbf{Yuichi Ota}: Conceptualization (equal); Data curation (lead); Formal analysis (lead); Methodology (lead); Investigation (lead);  Writing – original draft (lead); Writing – review \& editing (lead).
\textbf{Kentaro Kaneko}: Data curation (equal); Formal analysis (equal); Investigation (equal); Writing – review \& editing (equal).
\textbf{Takeyoshi Onuma}: Data curation (equal); Formal analysis (equal); Investigation (equal); Writing – review \& editing (equal).
\textbf{Shizuo Fujita}: Conceptualization (lead); Funding acquisition (lead); Project administration (lead); Supervision (lead); Validation (lead); Writing – review \& editing (equal).

\section*{DATA AVAILABILITY}
The data that supports the findings of this study are available within the article and its supplementary material.

\bibliography{JCP_ota}
\bibliographystyle{apsrev4-1}



%
%
\newpage

%



\end{document}